\documentclass[%
reprint,
superscriptaddress,
amsmath,amssymb,
aps,
]
{revtex4-2}
\usepackage{hyperref}
\hypersetup{
	colorlinks = true,
	linkcolor = black,
	anchorcolor = black,
	citecolor = black,
	filecolor = black,
	urlcolor = black
}
\newcommand{\RNum}[1]{\uppercase\expandafter{\romannumeral #1\relax}}
\usepackage{float}
\usepackage{filecontents}
\usepackage[english]{babel}
\usepackage{ragged2e}
\usepackage{caption}
\usepackage{ mathrsfs }
\DeclareCaptionFormat{myformat}{\fontsize{8}{6}\selectfont#1#2#3}
\usepackage{bm}
\usepackage{mhchem}
\usepackage{amsmath}
\usepackage{braket}
\usepackage{siunitx}
\usepackage{nicefrac,xfrac}
\usepackage{caption}
\usepackage{subcaption}
\usepackage{graphicx}
\usepackage{grffile}
\usepackage{changes}
\usepackage{graphicx}
\usepackage{dcolumn}
\usepackage{bm}
\usepackage{setspace}

\newcommand{\RN}[1]{\uppercase\expandafter{\romannumeral#1}}

\newcommand*{\hatH}{\hat{\mathcal{H}}}
\begin{document}
\title{Periodic quantum Rabi model with cold atoms at deep strong coupling }
\author{Geram Hunanyan}
\affiliation{Institut f{\"u}r Angewandte Physik, Universit{\"a}t Bonn, Wegelerstr. 8, 53115 Bonn, Germany}
\email{hunanyan@iap.uni-bonn.de}

\author{Johannes Koch}
\affiliation{Institut f{\"u}r Angewandte Physik, Universit{\"a}t Bonn, Wegelerstr. 8, 53115 Bonn, Germany}

\author{Stefanie Moll}
\affiliation{Institut f{\"u}r Angewandte Physik, Universit{\"a}t Bonn, Wegelerstr. 8, 53115 Bonn, Germany}

\author{Enrique Rico} 
\affiliation{Department of Physical Chemistry, University of the Basque Country UPV/EHU, Box 644, 48080 Bilbao, Spain}
\affiliation{Donostia International Physics Center, 20018 Donostia-San Sebastián, Spain}
\affiliation{EHU Quantum Center, University of the Basque Country UPV/EHU, P.O. Box 644, 48080 Bilbao, Spain}
\affiliation{IKERBASQUE, Basque Foundation for Science, Plaza Euskadi 5, 48009 Bilbao, Spain}

\author{Enrique Solano}
\affiliation{Kipu Quantum, Greifswalder Straße 226, 10405 Berlin, Germany}

\author{Martin Weitz}
\affiliation{Institut f{\"u}r Angewandte Physik, Universit{\"a}t Bonn, Wegelerstr. 8, 53115 Bonn, Germany}

\date{\today}
	\begin{abstract}
		The quantum Rabi model describes the coupling of a two-state system to a bosonic field mode. Recent theoretical work has pointed out that a generalized periodic version of this model, which maps onto Hamiltonians applicable in superconducting qubit settings, can be quantum simulated with cold trapped atoms. Here, we experimentally demonstrate atomic dynamics predicted by the periodic quantum Rabi model far in the deep strong coupling regime. The two-state system is represented by two Bloch bands of cold atoms in an optical lattice, and the bosonic mode by oscillations in a superimposed optical dipole trap potential. The observed dynamics beyond the usual quantum Rabi physics becomes relevant when the edge of the Brillouin zone is reached, and evidence for collapse and revival of the initial state is revealed at extreme coupling conditions. 

	\end{abstract}
\maketitle
\section{Introduction}
\vspace{-0.4cm}
The interaction of a two-state system with an oscillatory mode, as in a fully quantized form described by the quantum Rabi model, is among the most fundamental problems of quantum optics \cite{Rabi1935,Braak2016}. Experimental work on the interaction of two-level systems with quantized field modes has been carried out with Rydberg atoms in microwave cavities before being carried over to the optical domain \cite{Harochebook}. The obtained experimental results, for which the coupling strength between atoms and the electromagnetic field was above the decoherence rate, corresponding to the so-called strong coupling regime, are described by the celebrated Jaynes-Cummings model, which predicts the emergence of hybrid matter-light eigenstates \cite{JCM1963,HAFFNER2008155,Diaz2019,Langford2017,Markovic2018,Ciuti2005,Casanova2010,Peropadre2010}. Other than the Jaynes-Cummings model, the quantum Rabi model (QRM) is valid for arbitrary coupling strengths, given that beyond the co-rotating it also accounts for the counter-rotating terms of the interaction Hamiltonian, which leads to counter-intuitive effects as that excitations can be created out of the vacuum. The quantum Rabi physics becomes relevant as the coupling strength $g$ becomes comparable or even exceeds the bosonic mode frequency $\omega$, with the regime $g/\omega\geq1$ being termed the deep strong coupling (DSC) regime. Experimentally, quantum Rabi physics has been studied with superconducting Josephson systems, metamaterials, ion trapping, and with cold atom settings \cite{Dareau2018,Lv2018,Yoshihara2016,Bayer2017,Cai2021}. In recent work, by encoding the two-level system in the occupation of Bloch bands of cold trapped atoms, we have demonstrated quantum Rabi dynamics far in the DSC regime at interaction times at which the dynamics remains within the first Brillouin zone \cite{Koch2023}.\\
Here, we report the observation of collapse and revival of quantum Rabi dynamics in the DSC regime with cold trapped atoms. We monitor the atomic evolution at long interaction times beyond the first Brillouin zone, with the two-level system being encoded in the Bloch band structure and the bosonic mode in the oscillatory motion of a superimposed optical dipole trapping potential. In the investigated regime, the fact that the qubit information is stored in the band structure becomes significant such that the predictions of the QRM and the periodic quantum Rabi model (pQRM) differ, as understood from that half of the phase space available to the QRM is mapped to the external atomic structure. The pQRM can be considered as a generalization of the QRM, where the growth of the bosonic excitations is truncated at an earlier stage, generating an echo-like periodic cycle with doubled temporal periodicity. It essentially acts as a different light-matter interaction, emerging in a unique manner from the experimental context of cold atoms in variable potentials. The achieved normalized coupling strength of $g/\omega\simeq6.5$, meaning that the coupling clearly dominates over all other relevant system energies, compares favourably to the parameters of state of the art works of $g/\omega\simeq1.9$ obtained in the phase space of superconducting fluxonium systems \cite{Pechenezhskiy2020}. In our work, both collapse and revival of the excitation number are observed, as well as phase dependence of the prepared Schrödinger cat-like states. We attribute our experimental data to give evidence for a quantum simulation of the periodic quantum Rabi model (pQRM) in an AMO physics system at deep strong coupling.
\begin{figure*}
			\centering
			\includegraphics[width=\textwidth]{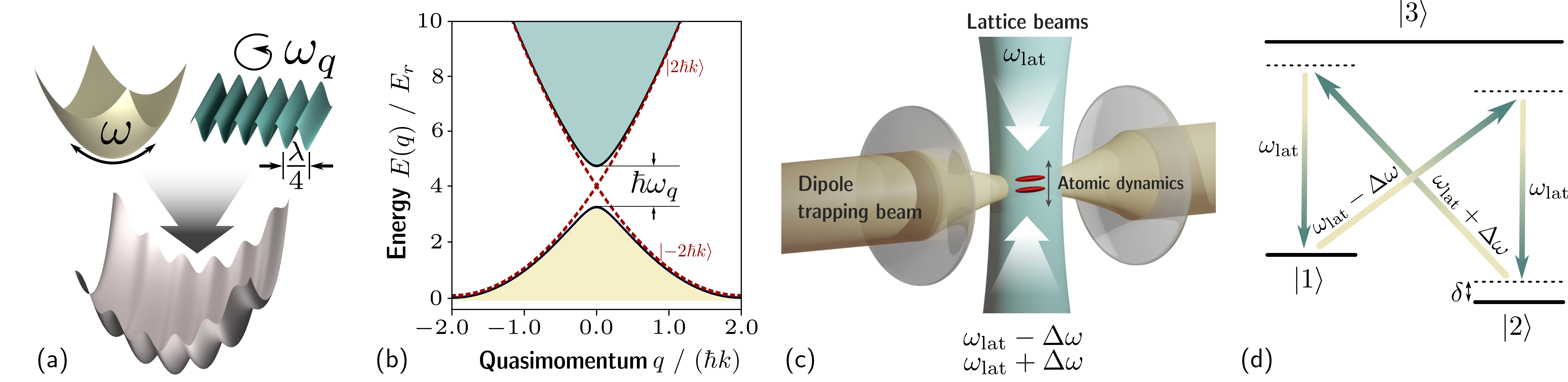}
			\caption{ \justifying (a) Cold rubidium atoms are subjected to the potential obtained by superimposing a harmonic trap (left) and a periodic lattice potential (right). The image also shows the corresponding oscillatory modes, an oscillation mode at frequency $\omega$ in the harmonic potential and a mode at frequency $\omega_q$ at the first band gap of the lattice. In the combined potential the coupling of modes can exceed their eigenfrequencies. (b) Atomic dispersion relation in a lattice of spatial periodicity $\lambda/4$. (c) Scheme of the setup. A near the centre harmonic confinement for cold rubidium atoms is realized with a focused dipole trapping laser beam. The additional high spatial periodicity lattice potential is realized by driving with an optical beam of frequency $\omega_\mathrm{lat}$ and two counterpropagating components with frequencies $\omega_\mathrm{lat}+\Delta\omega$ and $\omega_\mathrm{lat}-\Delta\omega$, see (d) for the coupling scheme.} 
			\label{fig:setup}
			\vspace{-2em}
		\end{figure*}\\
\section{Theoretical Description}
\vspace{-0.4cm}
The basic principle of the scheme realised in our work is shown in Fig. \Ref{fig:setup}. Sub-figure (a) illustrates the two quantum mechanical oscillation modes relevant here, which in the first case are generated by the oscillation of atoms in a harmonic trap potential, in the second case by Bragg reflection in a lattice potential from the splitting of the two lowest Bloch bands, the latter realising a two-state system. The superposition of the two potentials leads to an extremely strong coupling of the two quantised atomic oscillation modes. In our implementation, the harmonic trap potential is generated by a focused optical dipole trapping beam and the superimposed lattice potential via the dispersion of multi-photon Raman-transition \cite{Ritt2006,Salger2007}, resulting in a spatial periodicity of $\lambda/4$, where $\lambda$ is the wavelength of the driving optical beams. Formally, the system can be described by the Hamiltonian
	\begin{equation}
		\hatH = \frac{\hat{p}^2}{2m}+\frac{m\omega^2}{2}\hat{x}^2+\frac{V}{2}\cos(4k\hat{x})
		\label{eq:pqrm_ham}
	\end{equation}
	where  $\hat{x}$ and $\hat{p}$ are the position and momentum operators, $m$ is the atomic mass, $V$ the lattice depth, $k=2\pi/\lambda$, and $\omega$ is the harmonic trap frequency. Fig. \Ref{fig:setup}(b) shows the dispersion relation of atoms in the lattice in a representation that is centred around the position of the first band crossing at $p=\pm2\hbar k$ as a function of the atomic quasi-momentum $q$. At the band crossing, i.e. at $q=0$, the eigenstates of the two-level system are given by $\ket{g}=\frac{1}{\sqrt{2}}(\ket{-2\hbar k}+\ket{+2\hbar k})$ and $\ket{e}=\frac{1}{\sqrt{2}}(\ket{-2\hbar k}-\ket{+2\hbar k})$, respectively. \\	
	We introduce an eigenbasis of the momentum operator $|k\rangle = |q,n_{b}\rangle$ such that the description of the momentum eigenvalue is split into a continuous part $q$ defined in the interval $(-2\hbar k, 2 \hbar k]$ and an integer part $n_{b} \in \mathcal{Z}$ that defines a band index. The wave-functions for these states are given by $\langle x| q, n_{b} \rangle = e^{iqx/\hbar} e^{-i2kx} e^{i4n_{b} kx} $. The position operator $\hat{x}$ can be represented by the derivative of the quasi-momentum $q$ within a band, i.e. $\hat{x} = -i \hbar \frac{\partial}{\partial q}$, but will also induce a coupling between the bands at the boundaries of the interval (see Appendix B).	
	Within a band, the Hamiltonian of the system is just given by the harmonic term, $\hat{q}^2/2m+(m/2)\omega^2\hat{x}^2 =\hbar\omega(\hat{a}^\dagger\hat{a}+1/2)$, where $\hat{a}^\dagger=\sqrt{\frac{m\omega}{2\hbar}}(\hat{x}-\frac{i}{m\omega}\hat{q})$ is the creation operator of the bosonic field mode. If we project onto the two lowest bands $n_b=\{0,1\}$, the structure of a two-level system (qubit) appears. Within the first Brillouin zone, the resulting quantum Rabi Hamiltonian can be written as 	
	\begin{equation}
			\hatH_\mathrm{QRM} = \hbar \omega \hat{a}^\dagger\hat{a}+\frac{\hbar \omega_\mathrm{q}}{2} \hat{\sigma}_z + i\hbar g \hat{\sigma}_x(\hat{a}^\dagger-\hat{a}) 
			\label{eq:qrm_ham} .
	\end{equation} 
	Here, $\hat{\sigma}_x$ and $\hat{\sigma}_z$ are Pauli matrices acting on coarse-grained wave-functions in upper and lower bands, respectively, while $\hbar \omega_q$ denotes the splitting between bands corresponding to the qubit spacing and with a coupling $g = k\sqrt{2\hbar\omega/m}$. Formally, the quantum Rabi Hamiltonian of Eq.~\eqref{eq:qrm_ham} is derived from Eq.~\eqref{eq:pqrm_ham} in the absence of an Umklapp term \cite{Felicetti2017}. The complete dynamics also has a boundary term (see Appendix B), which introduces the notion of a periodic quantum Rabi Model (pQRM) and it is the subject of this manuscript. As we will see below, already at an early stage of the evolution, exactly at $\omega t / 2 \pi = 0.25$, the wave function of the collective system experiences an echo-like return such that the collapse and revival dynamics is modified.
	Interestingly, characteristic signatures of the pQRM, such as a modified pattern of collapse and revival compared to the QRM, can be described analytically by using a perturbative approach in a position-momentum $(x,p)$ phase space picture, as shown in the Appendix D.\\		
	Given that the two-level system in the used quantum Rabi implementation is stored in the band structure, effects beyond usual quantum Rabi physics can arise when one reaches the edge of the first Brillouin zone. For the here used lattice with spatial periodicity $\lambda/4$ the relation between momentum and quasi-momentum for the first two bands, see also Fig. \Ref{fig:setup}(b), is $q=p-2\hbar k$ for $p\geq 0$ and $q=p+2\hbar k$ for $p<0$ respectively, such that the quasi-momentum is restricted to $q\in(-2\hbar k, 2\hbar k]$. Essentially, storage of the qubit in the band structure itself introduces a folding in the Bloch band structure. This results in collapse and revival effects that are distinctly modified with respect or predictions of the original QRM.
	\section{Experimental Setup and Procedure}
	\vspace{-0.4cm}			
	Our setup, see also the schematics of Fig. \Ref{fig:setup}(c), is a modified version of an apparatus used in earlier works \cite{Salger2007,Ritt2006,Koch2023}. Initially a Bose-Einstein condensate of rubidium atoms (\textsuperscript{87}Rb) in the $m_F=-1$ spin projection of the $F=1$ ground state is produced in the quasi-static optical dipole trapping potential imprinted by a focused beam emitted by a CO$_2$-laser operating near $\SI{10.6}{\micro\meter}$ wavelength. The beam power $P$ is then adiabatically increased to reach a desired value of the trapping frequency $\omega\propto\sqrt{P}$ for quantum Rabi manipulation. Atoms are in addition exposed to a high spatial ($\lambda/4$) periodicity lattice potential, where $\lambda =\SI{783.5}{nm}$ denotes the wavelength of the driving laser beams. The potential of corresponding periodicity is generated by off-resonantly driving four-photon Raman transitions between the $m_F=-1$ and $m_F=0$ ground state sub-levels of $F=1$ over the $5P_{3/2}$ excited sate manifold using a beam of frequency $\omega_\mathrm{lat}$ and two superimposed counter propagating beams of frequencies $\omega_\mathrm{lat}+\Delta\omega$ and $\omega_\mathrm{lat}-\Delta\omega$ (Fig. \Ref{fig:setup}(d)). Following the adiabatic intensity ramp of the dipole trapping beam, atoms are prepared at the position of the first band crossing of the high spatial periodicity lattice, see also the dispersion relation of Fig. \Ref{fig:setup}(b), by means of Bragg diffraction. After subsequent activation of the lattice beams, atoms are exposed to the combined potential as indicated in Fig. \Ref{fig:setup}(a). Typical experimental parameters are harmonic trapping frequencies $\omega/2\pi \in [350, 750]$Hz, resulting in normalized coupling $g/\omega \in [4.1, 6.5]$, which is far in the DSC regime. The investigated regime for the two-level qubit splitting is $\omega_q/2\pi \in [0,5.5]$kHz.\\	
	At the end of the atom manipulation phase, both the lattice beams and the dipole trapping beam are extinguished, after which absorption imaging is employed for detection. In the course of the measurements, data was recorded analysing the real-space distribution, as probed by imaging directly after manipulation, as well as recording of the momentum distributions, for which time-of-flight imaging was used. For the former measurements, except when recording mean displacements, data analysis was performed after deconvolution with the determined point spread function of the imaging system, as to reduce systematic effects stemming from the $\simeq\SI{6.5}{\micro\meter}$ instrumental resolution of our imaging system. For the present measurements investigating long interaction times of quantum Rabi manipulation, relatively low atom numbers ($\sim800$) are used, to reduce interaction effects.
\begin{figure}
	\centering
	\includegraphics[width=1\linewidth]{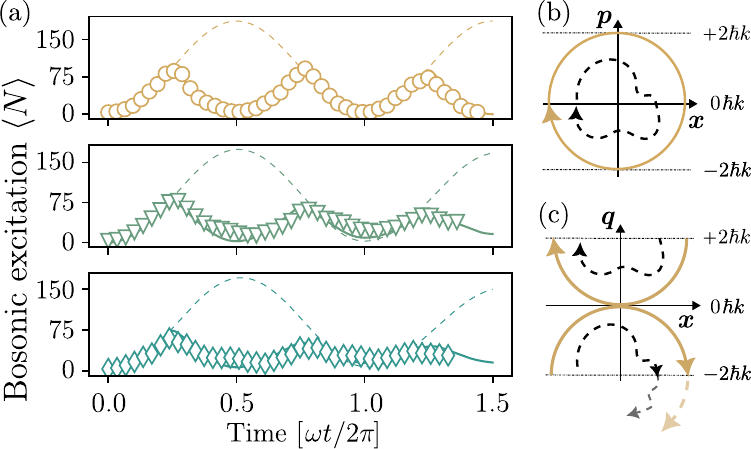}
	
	\caption{\justifying (a) Temporal evolution of the number of excitations $\langle N\rangle$ for a qubit splitting $\omega_q/2\pi\rightarrow0$ (top), $\SI{800(14)}{\hertz}$ (middle), and $\SI{1280(21)}{\hertz}$ (bottom), and a harmonic trapping frequency of $\omega/2\pi=\SI{346(8)}{\hertz}$ (relative coupling $g/\omega\simeq6.53)$. Atoms were initially prepared at a momentum of $p=-2\hbar k$. The solid lines are theory predictions based on the pQRM, and the dashed line for the (usual) QRM. (b) Illustration of the temporal evolution of atomic wavepackets in phase space ($x$: position, $p$: momentum) for $\omega_q=0$ (yellow solid line) and $\omega_q>0$ (black dashed line) respectively. (c) As in (b), now plotted in a position ($x$) - quasimomentum ($q$) representation. The transparent dashed lines indicate corresponding trajectories predicted in the (usual) QRM, for which no remapping of $q$ onto the Brillouin zone of the lattice occurs.}
	
	\vspace{-2em} 
	\label{fig:bosonic_excitation}
\end{figure}
\section{Results and discussion}
\vspace{-0.4cm}
To begin with, we have investigated the temporal evolution of the bosonic excitation number $\langle N\rangle$, with: $\hbar\omega \left( \langle N \rangle +\frac{1}{2}\right)  = \frac{m\omega^2}{2}\langle x^2 \rangle + \frac{1}{2m}\langle q^2 \rangle$, for times up to beyond the expected revival. Atoms were initially prepared at the trap centre with a momentum of $p=-2\hbar k$, for which the quasimomentum $q$ vanishes.  The data points in Fig.~\Ref{fig:bosonic_excitation}(a) give the temporal variation of the mean excitation number, as derived from the rms spread of the experimental in situ and time-of-flight imaging data, versus the interaction time of quantum Rabi manipulation, for different lattice depths $V=2\hbar\omega_q$ (top to bottom: $\omega_q/2\pi\rightarrow0$, $\SI{800(14)}{\hertz}$, and $\SI{1280(21)}{\hertz}$).  We observe a periodic pattern of the excitation number oscillating with half temporal period of the harmonic potential $T=2\pi/\omega$, which for larger qubit spacings reduces in magnitude, creating the pQRM. The experimental data is in good agreement with theoretical pQRM predictions, see Eq.~\eqref{eq:pqrm_ham} with the described identifications between $p$ and $q$ as to keep the quasimomentum $q$ in the first Brillouin zone. However, the observed physics does not follow the QRM predictions of Eq.~\eqref{eq:qrm_ham}, see the solid and dashed lines, respectively. Figs.~\Ref{fig:bosonic_excitation}(b) and \Ref{fig:bosonic_excitation}(c) qualitatively illustrate the expected atomic dynamics in position ($x$)-momentum- ($p$) and position ($x$)-quasimomentum- ($q$) phase space representations respectively (see also SM for corresponding experimental measurements). Here, the yellow solid line gives the expected variation for the trivial case of $\omega_q=0$, which corresponds to a usual (shifted) harmonic oscillator dynamics in position-momentum (position-quasimomentum) space, respectively, and the dashed line illustrates an example for the non-trivial case of $\omega_q>0$. The observed temporal variations of the mean excitation number (Fig.~\Ref{fig:bosonic_excitation}(a)), given it being proportional to the rms distance from the origin in position quasimomentum space, of half the harmonic oscillator period $T=2\pi/\omega$, is well understood from the corresponding trajectories at least for not too large values of $\omega_q$. For comparison, the dashed lines in Fig.~\Ref{fig:bosonic_excitation}(c) illustrate the expected behaviour predicted in the (original) quantum Rabi model (Eq.~\eqref{eq:qrm_ham}), for which the periodicity equals the full harmonic oscillator cycle.\\
  \begin{figure}
  	\centering
  \includegraphics[width=1\linewidth]{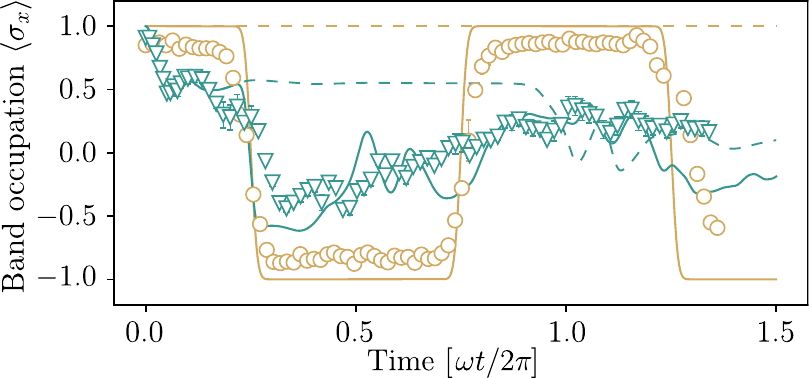}
  	\caption{ \justifying Temporal variation of the average value of the Bloch band occupation $\langle \sigma_x\rangle$ for $\omega/2\pi=\SI{346(8)}{\hertz}$ and qubit spacings of $\omega_q/2\pi\rightarrow0$ (yellow circles) and $\SI{1750(25)}{\hertz}$ (green triangles). The solid and dashed lines are theory predictions for the pQRM and the usual QRM, respectively. }
  	\label{fig:sigmax_all}
  	  	\vspace{-2em} 
  \end{figure}
  To study the effects of the band mapping in more detail, we have analysed the temporal variation of the mean Bloch band occupation $\langle \sigma_x\rangle$, which can be expressed in the basis of the band eigenstate numbers $\hat{\sigma}_x=\ket{n_b=0}\bra{n_b=0}-\ket{n_b=1}\bra{n_b=1}$, with $n_b=0,1$ for $p=q-2\hbar k$ and $p=q+2\hbar k$ respectively. Corresponding experimental data is shown in Fig. \Ref{fig:sigmax_all} for different qubit splittings. While at small lattice depth the expectation value of the band index remains constant until the edge of the Brillouin zone is reached and the bands are remapped, at higher lattice depth the modulus of $\langle \sigma_x\rangle$ reduces and oscillations are observed (visible most clearly near times $t=0$ and $2\pi/\omega$), as attributed to the Rabi oscillations between the momentum states $\pm2\hbar k$ respectively. The oscillations are suppressed at smaller values of $\omega_q$, since then the coupling term, which is proportional to $\hat{\sigma}_x$ \cite{Wolf2012}, dominates over all other energy scales, and appear only for larger values of the qubit splitting, upon which the dispersive DSC regime is reached. Corresponding behaviour has for small interaction times also been observed in earlier work of our groups \cite{Koch2023}, and the present results generalize these observations to beyond the first Brillouin zone.\\ 
  	\begin{figure}[h]
  	\centering
  	\includegraphics[width=1\linewidth]{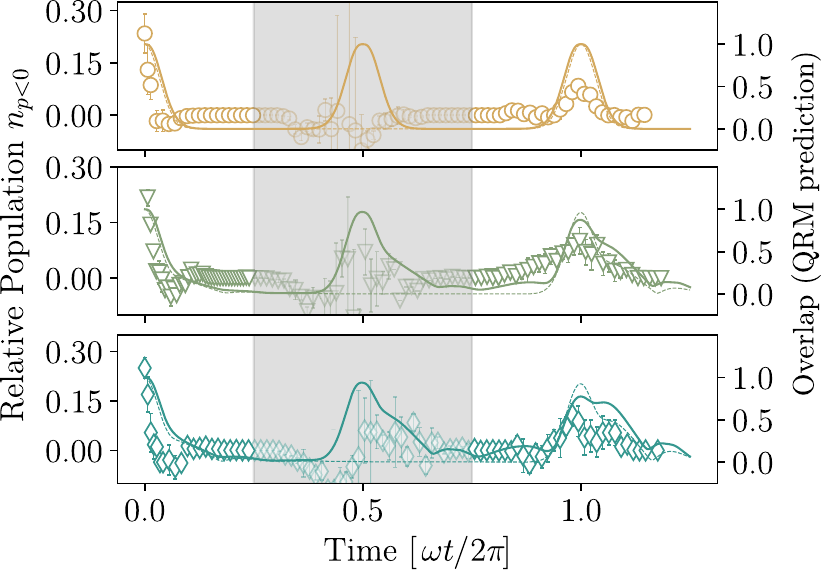}

  	\caption{ \justifying Collapse and revival of an initial state. The data points give relative number of atoms $n_{p < 0}=(N_{p<0} – N_{p>0}) /(N_{p<0} + N_{p>0}$), with $N_{p<0}$ ($N_{p>0}$) as the atom number detected with momentum $p<0$ ($p>0$), versus the interaction time for $\omega/2\pi=\SI{650(21)}{\hertz}$ and qubit splittings $\omega_q/2\pi\rightarrow0$ (top), $\SI{800(14)}{\hertz}$ (middle), and $\SI{1280(21)}{\hertz}$ (bottom). The solid lines are theory for the pQRM, and the dashed lines give the overlap with the initial state (see scale on the right hand side) as predicted in the (usual) QRM. The gray shaded area indicate regions where the experimental signal is affected from large instrumental phase fluctuations.}
  	\label{fig:cnr}
  \end{figure}  
  In subsequent measurements, we have prepared atomic wavepackets in the qubit eigenstates of the system, which are superpositions of momentum picture eigenstates. For this, atoms were irradiated by two simultaneously performed Bragg pulses of counterpropagating momentum transfer, such that depending on the relative phase different qubit initial states can be prepared. Fig. \Ref{fig:cnr} gives data investigating collapse and revival of an initially prepared qubit eigenstate $\ket{g}=\frac{1}{\sqrt{2}}(\ket{-2\hbar k}+\ket{+2\hbar k})$ for a harmonic trap frequency $\omega/2\pi\simeq\SI{650}{Hz}$ ($g/\omega\simeq4.8$) and different values of the lattice depth. For the measurement, after a variable interaction time in the combined potential, a $\pi/2$ four-photon Raman pulse tuned to drive transfer between the momentum states $-2\hbar k$ and $2\hbar k$ was applied such that when the initial state is fully revived, atoms are transferred to the momentum state $-2\hbar k$ and we have $\sigma_z =1$. The vertical scale of the plots in Fig. \Ref{fig:cnr} shows the relative number of atoms observed with a negative momentum ($p<0$) in the time of flight images with respect to the total atom number, which constitutes a measurement of $\langle \sigma_z \rangle$. The top plot, corresponds to a vanishing lattice depth ($\omega_q\rightarrow0$) such that this experiment realizes a trapped atom interferometer, shows a revival at a full oscillation time of $t=2\pi/\omega$. The middle and lower panels, as recorded for increased lattice depth, show revivals with visible substructures. The experimental data qualitatively agrees with predictions based on the periodic quantum Rabi model (solid lines), and we attribute the reduced contrast of the revival signal mainly to the finite atomic velocity distribution. The shaded area at near half the revival time corresponds to a region where large phase fluctuations, attributed to mechanical vibrations of the Raman beams with respect to the dipole trapping beam, become relevant given the here reversed propagation direction of atomic wavepackets paths with respect to preparation, and we consider this region as inaccessible to the experiment. For comparison, the dashed line gives the expected overlap of the initial state predicted in the (standard) quantum Rabi model, for which no revival at half the oscillation time is expected.
	\begin{figure}[]
		\centering
		\includegraphics[width=1\linewidth]{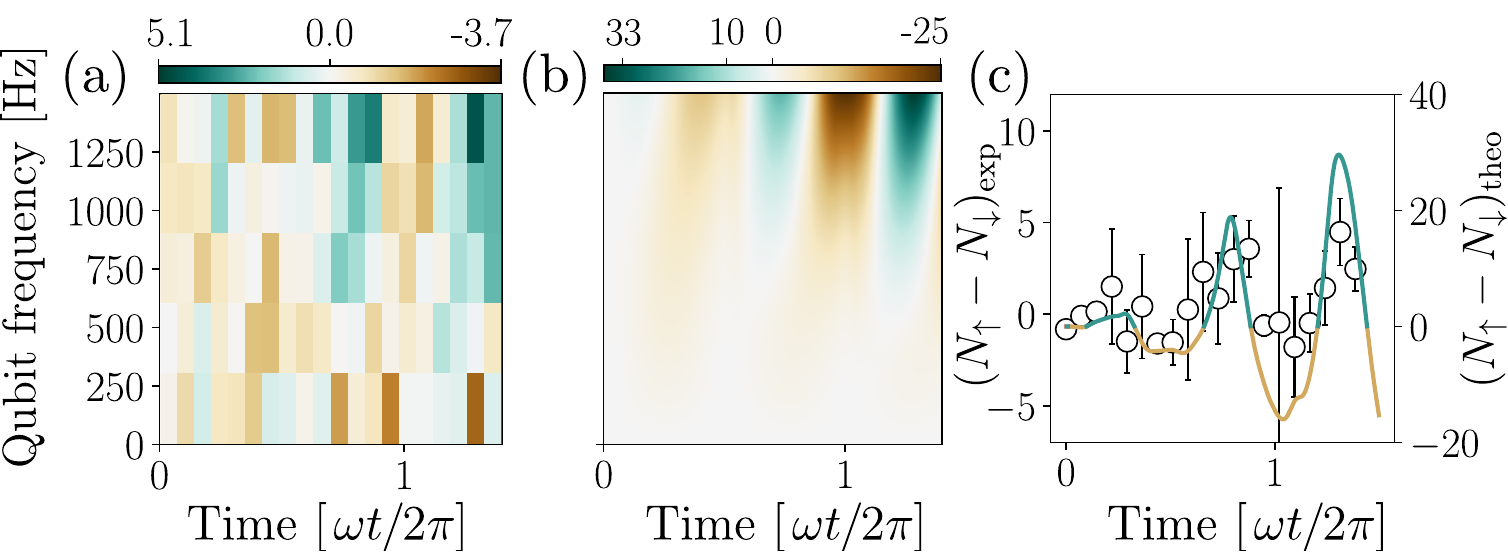}

		\caption{ \justifying (a) Difference ($N_\uparrow-N_\downarrow)$ of observed mean excitation numbers between when initially preparing atoms in upper and lower qubit states $N_\uparrow$ and $N_\downarrow$ versus both the interaction time and the qubit spacing $\omega_q/2\pi$, in a colour-code representation for $\omega/2\pi\simeq\SI{350}{\hertz}$, $g/\omega\simeq6.53$. (b) Corresponding theory. (c) Data for $\omega_q/2\pi=\SI{1250(23)}{\hertz}$ exemplary drawn as a diagram, along with theory (solid line). }
		\label{fig:ndiff}
		  	\vspace{-2em} 
	\end{figure} 

To study the dependence of pQRM evolution on the qubit state encoded in the Bloch band structure, we have analysed the number of created system excitations both when initially preparing atoms in eigenstates $\ket{g}=\frac{1}{\sqrt{2}}(\ket{-2\hbar k}+\ket{+2\hbar k})$ and $\ket{e}=\frac{1}{\sqrt{2}}(\ket{-2\hbar k}-\ket{+2\hbar k})$. Fig. \Ref{fig:ndiff}(a) gives experimental data for the difference in the correspondingly obtained excitations versus time for different lattice depths. We observe a clear difference in the number of created excitations for two different used relative phases at qubit splitting above near $\omega_q/2\pi\simeq\SI{700}{\hertz}$. This is in agreement with theory (Fig. \Ref{fig:ndiff}(b)), though in the experiment again the contrast is reduced. As an example, Fig. \Ref{fig:ndiff}(c) gives a plot of the temporal variation of the difference in observed excitations for $\omega_q/2\pi\simeq\SI{1250}{\hertz}$. Formally, at a small qubit splitting the oscillating wavepackets can well be described as Schrödinger cat states, while they become highly entangled states at larger values of the qubit splitting. The agreement of the experimental data with the theory is attributed as evidence that coherence is maintained in the dispersive DSC regime of the pQRM.
\section*{Conclusions}
\vspace{-0.4cm}
To conclude, we have observed collapse and revival effects of the dynamics in a cold atom based quantum simulation of the quantum Rabi physics at extreme parameter regimes. Our experimental data is in good agreement with theory based on a generalized, periodic variant of the usual QRM, the physical origin being the periodic nature of the atomic Brillouin zone of cold atoms in a lattice.\\
For the future, it would be interesting to generalize the reported observations to using atoms with tunable interactions using Feshbach tuning (e.g. \textsuperscript{85}Rb or \textsuperscript{39}K), such that both the limits of negligible and stronger interactions can be explored. Other perspectives, inspired also by the formal analogy of the system Hamiltonian to superconducting qubit systems, include quantum information processing applications \cite{Manucharyan2009,Lamata2018,Hwang2015}, as well as the search for novel quantum phase transitions \cite{Heyl2018,Yang2023}. 
\section*{Acknowledgements}
\vspace{-0.4cm}
We acknowledge support by the DFG within the project We 1748-24 (642478), the focused research center SFB/TR 185 (277625399) and the Cluster of Excellence ML4Q (390534769). E.R. is supported by the grant PID2021-126273NB-I00 funded by MCIN/AEI/ 10.13039/501100011033 and by "ERDF A way of making Europe" and the Basque Government through Grant No. IT1470-22. This work was supported by the EU via QuantERA project T-NiSQ grant PCI2022-132984 funded by MCIN/AEI/10.13039/501100011033 and by the European Union ``NextGenerationEU''. We want to thank Razmik Unanyan for his contributions to the perturbative treatment of the deep strong coupling regime given in the Appendix D of the Appendix. 
\section*{Appendix}
\vspace{-0.4cm}
\subsection{Additional Data and Experimental Details}
\vspace{-0.4cm}
The near harmonic trapping potential for the cold cloud of rubidium atoms is generated with a focused laser beam ($\SI{46}{\micro\meter}$ diameter) derived from a CO$_2$-laser operating at near $\SI{10.6}{\micro\meter}$ wavelength. Despite the large detuning of the mid-infrared radiation from the lowest electronic resonances (the rubidium D-lines) which results in a very low heating rate from spontaneous scattering, the quasi-static atomic polarizability leads to a confining potential. Both the optical lattice and the optical Bragg pulses are generated using optical radiation derived from a high power diode laser operating near $\lambda\simeq\SI{783.5}{nm}$ wavelength, which is detuned by approximately $\SI{3.3}{nm}$ to the red of the rubidium D2-line. The laser emission is split into two, and each of the partial beams pass an acousto-optic modulator (AOM) used to imprint different optical frequency components and then coupled into optical fibers and guided to the vacuum chamber in which the cold atom experiment takes place. The optical lattice potential, which has a spatial periodicity of $\lambda/4$, is generated by the dispersion of Doppler-sensitive Raman transitions \cite{Ritt2006}. For this, see also Figs. 1c and 1d of the main text, atoms are irradiated with two co-propagating beams of frequencies $\omega_\mathrm{lat}+\Delta\omega$ and $\omega_\mathrm{lat}+\Delta\omega$  and one counter-propagating beam of frequency $\omega$. Here, the $m_\mathrm{F}=-1$ and the $m_\mathrm{F}=0$ spin projections of $F=1$ are the ground states of the three-level scheme used, while the $5\textsuperscript{2}P_{3/2}$ manifold serves as the electronically excited state. We use a frequency offset between counter-propagating beams of $\Delta\omega/{2\pi}\simeq\SI{945}{kHz}$, which is large enough to suppress unwanted standing wave two-photon processes. A magnetic bias field of  $B=\SI{1.7}{G}$ removes the degeneracy of the Zeeman sublevels, and the used value of the two-photon detuning (see also Fig. 1d) is $\delta/{2\pi}\simeq\SI{210}{kHz}$. The optical fields exchange momentum with the atoms in units of four photon recoils, which is a factor two above the corresponding process in a usual standing wave lattice, resulting in a $\lambda/4$ spatial periodicity of the generated lattice potential.
\begin{figure}[h]
	\centering
	\includegraphics[width=1\linewidth]{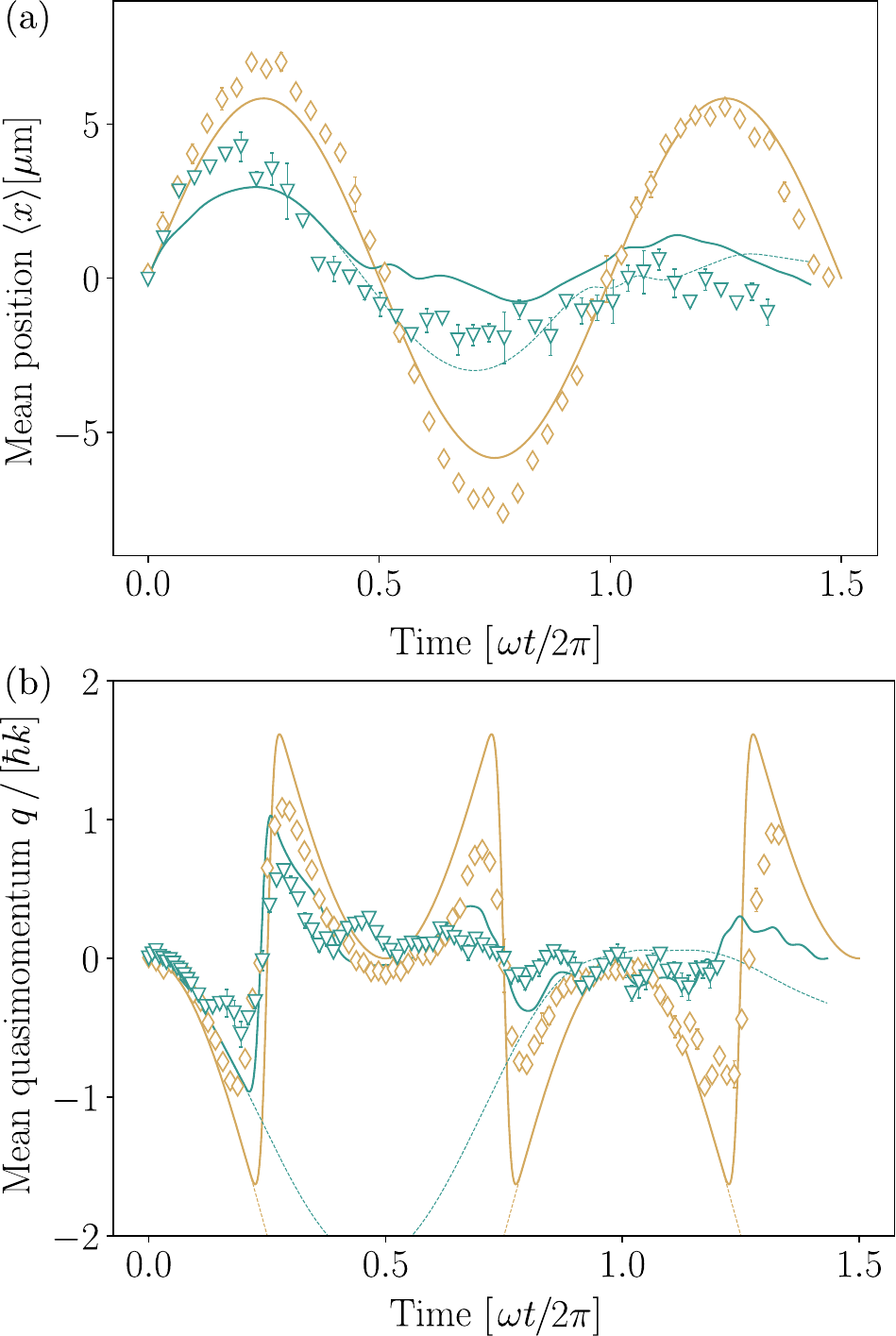}
	\caption{\justifying Temporal evolution of the measured values for the mean atomic position $\langle x\rangle$ (a) and mean quasimomentum $\langle q\rangle$ (b) for $\omega_q/2\pi\rightarrow0$ (yellow data points) and  $\omega_q/2\pi=\SI{1750(25)}{\hertz}$ (green data points), along with theory based on the perodic quantum Rabi model (solid lines) and the (usual) quantum Rabi model (dashed lines). Experimental parameters are as in Fig. 3 of the main text. }
	\label{fig:S1}
\end{figure} 
\twocolumngrid
We next give additional experimental data regarding the measurement shown in Fig. 3 of the main text, showing the temporal evolution of the Bloch band occupation in the combined lattice and harmonic trap for quantum Rabi manipulation. Figure \Ref{fig:S1} gives the observed corresponding temporal evolution of mean values of the atomic position $\langle x\rangle$, and quasi-momentum $\langle q\rangle$. As described in the main text, atoms here were initially prepared at a momentum of $p=-2\hbar k$ and in the trap center, the used bosonic mode frequency is $\omega/2\pi=\SI{346(8)}{\hertz}$, and the qubit spacing was $\omega_q/2\pi\rightarrow0$ (yellow circles) and \SI{1750(25)}{\hertz} (green triangles). In all cases, the experimental data well compares with predictions based on the periodic quantum Rabi model. Note that the experimental resolution of the imaging system ($\SI{6.5}{\micro\meter}$) is comparable to the trapped atomic cloud size, which limits the significance of a detailed analysis of the real space data. Nevertheless, the measurements allow us to qualitatively validate the illustrations for the expected atomic wavepacket trajectories indicated in Figs. 2b and 2c of the main text.
\begin{figure*}
	\centering
	\includegraphics[width=1\linewidth]{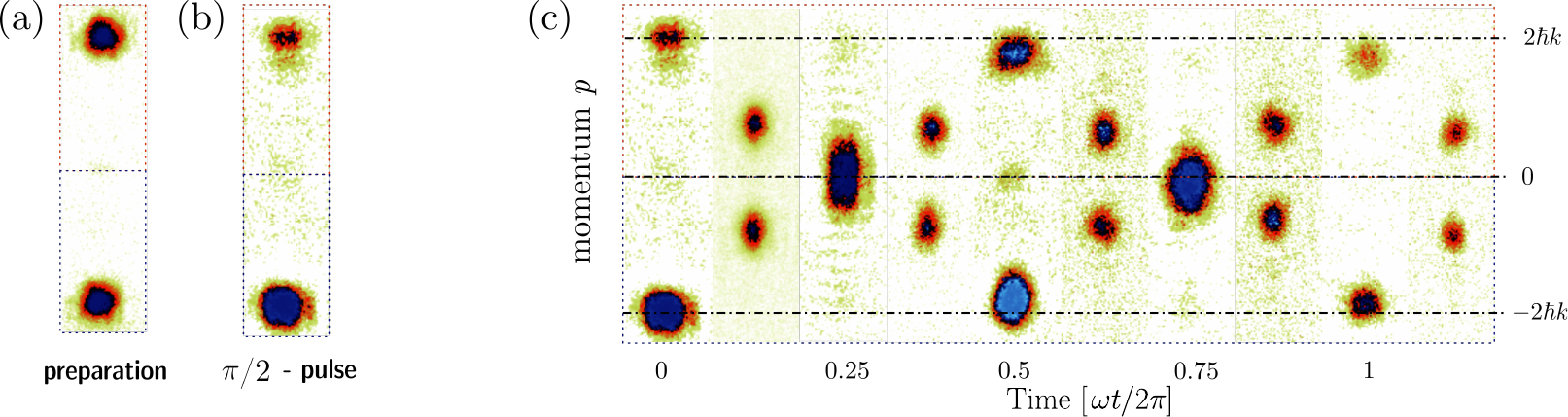}
	\caption{\justifying (a) Time-of light image analyzing the atomic momentum distribution following preparation of atoms in the qubit state $\ket{g}=\frac{1}{\sqrt2}(\ket{-2\hbar k}+\ket{2\hbar k})$. (b) Image recorded following an additional $\pi/2$ four-photon Raman pulse. (c) Series of time-of flight images recorded 
		with an additional, variable delay between the preparation pulse and the final $\pi/2$ readout pulse. }
	\label{fig:S2}
\end{figure*}  
Fig. \Ref{fig:S2} gives examples of obtained time-of-flight imaging data, as employed to evaluate the atomic momentum distribution at the end of quantum Rabi manipulation in the combined lattice and harmonic trap potential. The used free expansion time is $\SI{8}{\milli\second}$, after which an absorption image was recorded on a sCMOS camera. To begin with, Fig. S2a shows an absorption image recorded to analyze the atomic velocity distribution directly after preparing atoms in the qubit state  $\ket{g}=\frac{1}{\sqrt2}(\ket{-2\hbar k}+\ket{2\hbar k})$, corresponding to a superposition of two counterpropagating momentum picture states, and Fig. S2b an image after in addition applying a $\pi/2$-four-photon Raman pulse, resulting in transfer of atoms predominantly to state $\ket{-2\hbar k}$. As described in the main text, the from the data visible finite transfer efficiency is attributed to the atomic velocity distribution. Next, Fig. S2c shows a series of time-of flight images recorded after different atomic interaction times in the harmonic trap potential, with again atoms initially prepared in the qubit state  $\ket{g}=\frac{1}{\sqrt2}(\ket{-2\hbar k}+\ket{2\hbar k})$ as a superposition of counterpropagating wavepackets, and after the interaction time in the harmonic trap applying a $\pi/2$-four-photon Raman pulse, as to provide exemplary raw data for the top plot ($\omega_q/2\pi=0$) of Fig. 4 of the main text. Both at near half the oscillation time and a full oscillation time ($T=2\pi/\omega$) an enhanced population for atoms at negative final momentum $p<0$ is observed, as understood from the rephasing of wavepackets. Given that this corresponding to a single realization of the experiment, phase fluctuations between the trapping and Raman beams do not affect the contrast, the presence of revivals at both half and a full oscillation time is well understood. Upon subsequent realizations of the experiment, only the data at a full revival remains phase stable.
\subsection{Theoretical Methods}
\label{mapping}
\vspace{-0.4cm}
\subsubsection{Periodic Term of the Hamiltonian}
\vspace{-0.4cm}
The system is composed of a cloud of ultra-cold atoms exposed to two laser-induced potentials: a periodic lattice and a harmonic trap. When the atom density is sufficiently low, interactions among the atoms are negligible, and the system can be described with a single-particle Hamiltonian, 
\begin{equation}
	\hat{H}= \frac{\hat{p}^{2}}{2m} + \frac{V}{2} \cos \left( 4 k \hat{x} \right) + \frac{m \omega^{2}}{2} \hat{x}
\end{equation}
where, $\hat{p}=-i\hbar \frac{\partial}{\partial x}$ and $\hat{x}$ are momentum and position of an atom of mass $m$, respectively. Here, $\omega$ is the angular frequency of the atom motion in the harmonic trap, while $V$ and $4k$ are the depth and wave-vector of the periodic potential, respectively. The periodic lattice is resulting from a four-photon interaction with a driving field of wave-vector $k$ .

We will assume that the harmonic trap is slowly varying on the length-scale of the periodic potential. Under this assumption, the most suitable basis is given by the Bloch functions $\langle x | \phi_{n} \left(x\right) \rangle = e^{iqx/\hbar} e^{-i2kx} e^{i4nkx}$, where the first Brillouin zone is defined $q \in ( -2\hbar k,2 \hbar k ]$ and the band index $n \in \mathbb{Z}$.

It is straightforward to see that the momentum operator is diagonal in the Bloch basis, while the periodic potential introduces a coupling between adjacent bands
\begin{multline}
	\frac{\hat{p}^{2}}{2m} + \frac{V}{2} \cos \left( 4 k \hat{x} \right) |q,n\rangle = \\ \frac{1}{2m} \left[ q + \left( 2n-1\right) 2 \hbar k \right]^{2}|q,n\rangle 	+ \frac{V}{4} \left( |q,n+1\rangle + |q,n-1\rangle \right)
\end{multline}

Assuming that the system dynamics is restricted to the two bands with lowest energy, the periodic part of the Hamiltonian can be rewritten in the Bloch basis as 
\begin{align}
	\hat{H}_p=\frac{q^{2}}{2m} + \frac{2 \hbar k}{m} \sigma_{z} q + \frac{V}{4} \sigma_{x}
\end{align}
Hence, the periodic potential allows to encode the two-level system in the lowest two bands of the Bloch band structure. If the dynamics is kept in the same band, the harmonic potential introduces an operator which can be expressed as $\hat{x} = - i \hbar \frac{\partial}{\partial q}$ in the Bloch basis. This allows us to define the quasi-momentum operator $\hat{q}$ and the position operator $\hat{x}$, which satisfy the usual commutation relation $[\hat{x} , \hat{p}]= i \hbar$. 

In this way, we can rewrite the Hamiltonian as
\begin{equation}
	\hat{H} = \frac{\hat{q}^{2}}{2m} + \frac{m \omega^{2}}{2} \hat{x}^{2} + \frac{2 \hbar k}{m} \sigma_{z} \hat{q}  + \frac{V}{4} \sigma_{x}.
\end{equation}

\subsubsection{Quadratic Potential in the Bloch Basis}
\vspace{-0.4cm}
Let us now discuss the quadratic term $\frac{m \omega^{2} }{2} \hat{x}^{2}$ in the main Hamiltonian. In the Bloch basis, we can write

\begin{equation}
	\langle \tilde{q} , \tilde{n} | \hat{x}^{2} | q , n \rangle = \int^{+\infty}_{-\infty} dx~ x^{2} e^{i\left[ 4 \left( n - \tilde{n} \right) k + \left( q - \tilde{q} \right)/\hbar \right] x}.
\end{equation}

Considering diagonal elements in the qubit Hilbert space, i.e., setting $\tilde{n} = n$, we have 

\begin{equation}
	\langle \tilde{q} , n | \hat{x}^{2} | q , n \rangle = \int^{+\infty}_{-\infty} dx~ x^{2} e^{i \left( q - \tilde{q} \right)x/\hbar } = - \hbar^{2}\langle \tilde{q} , n |\frac{\partial^{2}}{\partial q ^{2}} | q , n \rangle.
\end{equation}

Hence, we see that the harmonic potential introduces an operator, diagonal in the qubit Hilbert space, which can be expressed as $\hat{r}=-i\hbar \frac{\partial}{\partial q}$, in the Bloch basis. This allows us to define the quasi-momentum operator $\hat{q}$ and the position operator $\hat{r}$, which satisfy the commutation relation $\left[ \hat{r} , \hat{q} \right] = i \hbar$. On the other hand, for $\tilde{n} = n$, the integral is different from zero only if $ 4 \hbar \left( n - \tilde{n} \right) k =  q - \tilde{q} $. Hence, the quadratic potential introduces a coupling between neighbouring bands, for states whose momenta satisfy $4 \hbar k =  q - \tilde{q} $, of the kind $\left( | 2 \hbar k, n_{b} \rangle \langle -2 \hbar k, n_{b} +1 | + \text{H.c.} \right)$. This effective coupling is due to the periodicity of the quasi-momentum, which mixes the bands $n=\{ 0 ,1 \}$ at the boundaries of the Brillouin zone. Such a coupling can be neglected as far as the system dynamics involves only values of the quasi-momentum $\hat{q}$ included within the first Brillouin zone.

\subsection{Mapping to Fluxonium Systems}
\label{Flux}
\vspace{-0.4cm}
The Fluxonium system is a circuit where we have in parallel a capacitor, an inductor and a Josephson junction, when the energies of each of the elements are, in comparison with one another, $E_J>E_C>E_L$. An analogy between the periodic quantum Rabi model presented in the main text and a superconducting fluxonium system \cite{Pechenezhskiy2020} can be shown in the following.
In principle, we only have one active node, \textit{a}. The equations of the flux going through this node are:
\begin{equation}
	C \ddot{\phi}_a=-\frac{\phi_a}{L}-J\sin\left(\frac{2\pi(\phi_a+ \Phi_{\text{clas}})}{\Phi_0}\right)
\end{equation}
where $\Phi_{\mathrm{clas}}$ is the external magnetic flux going through the spire defined by the Josephson junction and the inductor. From these equations we can propose the following Lagrangian:
\begin{equation}
	\mathscr{L}_{\mathrm{Flux}}=\frac{C}{2}\dot{\Phi}_a^2-\frac{\Phi_a^2}{2L}+\frac{J\Phi_0}{2\pi}\cos\left(\frac{2\pi(\phi_a+ \Phi_{\mathrm{clas}})}{\Phi_0}\right)
\end{equation}
With this we are already in situation of obtaining the Hamiltonian:
\begin{equation}
	H=\frac{q_a^2}{2C}+\frac{\Phi_a^2}{2L}-\frac{J\Phi_0}{2\pi}\cos\left(\frac{2\pi(\phi_a+ \Phi_{\text{clas}})}{\Phi_0}\right)
\end{equation}
We will rewrite it as follows:
\begin{equation}
	H=4E_C n_a^2+\frac{1}{2}E_L\Phi_a^2-E_J\cos\left(\frac{2\pi(\phi_a+ \Phi_{\mathrm{clas}})}{\Phi_0}\right)
\end{equation}
where we have defined $E_C=2e/8C$, $n=q/2e$, $E_L=1/L$ and $E_J=\frac{J\Phi_0}{2\pi}$. After this we will have to quantize the system, therefore getting:
\begin{equation}
	\hat{H}=4E_C \hat{n}_a^2+\frac{1}{2}E_L\hat{\Phi}_a^2-E_J\cos\left(\frac{2\pi(\hat{\phi}_a+ \Phi_{\text{clas}})}{\Phi_0}\right)
\end{equation}
This is as if we had a particle with mass inversely proportional to $E_C$ in the potential $V(\hat{\phi})=\frac{1}{2}E_L\hat{\Phi}_a^2-E_J\cos\left(\frac{2\pi(\hat{\phi}_a+ \Phi_{\text{clas}})}{\Phi_0}\right)$. Bear in mind that this potential depends on the $\Phi_{\mathrm{clas}}$ parameter, and that, tuning it, we can achieve different potentials which will lead to significantly different systems.
Explicitly, if we substitute of $\hat{\Phi}_a=4k\hat{x}$, $E_C=2\frac{k^2}{m}$, $E_J=\omega_q$, $E_L=\frac{m \omega^2}{16 k^2}$, $g=\left( 8 E_L E^3_C\right)^{1/4}$ and $\Phi_{\mathrm{class}} = \pi$, we arrive at an exact mapping between the atomic physics model and the superconducting circuit model. Comparing the energy scales given in \cite{Pechenezhskiy2020} to the parameters used in our setup yields a relative coupling strength of $g/\omega\approx1.91$ and a ratio between the qubit splitting and bosonic mode $\omega_q/\omega\approx2.42$.

\subsection{Perturbative Deep Strong Coupling Regime: Analytical Treatment}
\label{pDSC}
\vspace{-0.4cm}
In this section, we present an analytical approach to derive the dynamics of the pQRM using perturbation theory. To begin, we evaluate the expectation value of the mean position $\left\langle x \right\rangle$, see also in Figure \ref{fig:S1} of the Appendix, employing a perturbative approach applied to the system Hamiltonian described by equation (1) of the main text. This perturbative analysis offers a new perspective on the system Hamiltonian, complementing the approach outlined in the main text via Bloch band mapping. 
In earlier works regarding the spectral classification of the quantum Rabi model, it has been shown that when increasing the relative coupling strength $g/\omega$, with $\omega$ being the bosonic mode frequency to the dominating energy in the system, one moves from the 'usual' deep strong coupling regime (DSC, $g/\omega\simeq1$) to the so called perturbative deep strong coupling regime (pDSC, $g/\omega\gg1$) \cite{Rosatto2017}. It is in this limit, where in contrast to the earlier discussed regimes, where the coupling strength can be assumed to be a perturbation of the system, now it being the dominating energy, while the normalized qubit frequency $\omega_{\mathrm{q}}/\omega$ is now regarded as the perturbation. Hence, by using perturbation theory predictions of observables of the pQRM in a real space description based on using the variables $x$ and $p$ can be given. To compare the effectiveness of this perturbative method with a straightforward numerical approach, we will also present the overlap and fidelity of the respective systems in the following. It is noteworthy to mention, that a similar perturbative approach can be used to derive the dynamics in the QRM far in the deep strong coupling regime using the parity symmetry of the system \cite{PhysRevLett.105.263603}.\\
The Hamiltonian of the model is (eq. (1) of the main text)%
\[
H=\frac{p^{2}}{2m}+\frac{m\omega^{2}}{2}x^{2}+\frac{V}{2}\cos\left(
4kx\right)
\]
In this note we use the following units: $m=\hbar=\omega=1$, then the Hamiltonian takes the form
\begin{equation}
	H=\frac{p^{2}}{2}+\frac{1}{2}x^{2}+\frac{V}{2}\cos\left(  4kx\right)
	\label{Hamiltonian_units}%
\end{equation}
By standard definition of 
\begin{align*}
	x  &  =\frac{1}{\sqrt{2}}\left(  a^{\dagger}+a\right)  ,\\
	p  &  =i\frac{1}{\sqrt{2}}\left(  a^{\dagger}-a\right)
\end{align*}
The Hamiltonian (\ref{Hamiltonian_units}) can be written as
\begin{equation}
	H=a^{\dagger}a+\frac{1}{2}+\frac{V}{2}\cos\left(  2g\left(  a^{\dagger
	}+a\right)  \right)  , \label{hamiltonian22}%
\end{equation}
where
\[
g=2k\sqrt{\frac{\hbar\omega}{2m}}=\sqrt{2}k.
\]
For purposes of a perturbative calculation it is convenient to choose the unperturbed wave functions as Fock states. 
We seek a solution of the Schrödinger equation
\begin{equation}
	i\frac{\partial}{\partial t}\left\vert \Psi\right\rangle =\left[  a^{\dagger
	}a+\frac{1}{2}+\frac{V}{2}\cos\left(  2g\left(  a^{\dagger}+a\right)  \right)
	\right]  \left\vert \Psi\right\rangle \label{schrodinger11}%
\end{equation}
in the form
\[
\left\vert \Psi\left(  t\right)  \right\rangle =%
{\displaystyle\sum\limits_{n=0}^{\infty}}
a_{n}\left(  t\right)  \left\vert n\right\rangle .
\]
Then $a_{n}\left(  t\right)  $ obey the equation
\begin{multline}	
	i\frac{\partial}{\partial t}a_{n}\left(  t\right)  = \\ \left(  n+\frac{1}%
	{2}\right)  a_{n}\left(  t\right)  +\frac{V}{2}%
	{\displaystyle\sum\limits_{m=0}^{\infty}}
	\left\langle m\right\vert \cos\left(  2g\left(  a^{\dagger}+a\right)  \right)
	\left\vert n\right\rangle a_{m}\left(  t\right)  .
\end{multline}
If the perturbation $\frac{V}{2}\cos\left(  2g\left(  a^{\dagger}+a\right) \right)  $ is small, we can replace the eigenenergy of the Hamiltonian in eq. \eqref{hamiltonian22}
\[
E_{n}\approx n+\frac{1}{2}+\frac{V}{2}\left\langle n\right\vert \cos\left(  2g\left(
a^{\dagger}+a\right)  \right)  \left\vert n\right\rangle
\]
and the eigenstates by the Fock state $\left\vert n\right\rangle $. For the validity of the perturbation theory the matrix elements
\[
\left\langle n\right\vert \cos\left(  2g\left(  a^{\dagger}+a\right)  \right)
\left\vert m\right\rangle
\]
must satisfy the condition%
\begin{equation}
	\frac{V}{2}\left\vert \left\langle m\right\vert \cos\left(  2\sqrt
	{2}gx\right)  \left\vert n\right\rangle \right\vert <<\left\vert
	n-m\right\vert \label{Perturbation_condition}%
\end{equation}
for any $\left\vert m\right\rangle $ and $\left\vert n\right\rangle $ Fock states. 
\\
The cosine matrix elements can be expressed in terms of generalized
Laguerre polynomials (see:   S. Gradshteyn and I. M. Ryzhik, Table of
Integrals, Series, and Products, 7th ed. Academic Press, San Diego, 2007
page 806). We will consider only $m=n+2l$
\begin{multline}
	\left\langle n\right\vert \cos\left(  2\sqrt{2}gx\right)  \left\vert
	n+2l\right\rangle =\\ 
	\sqrt{\frac{n!}{\left(  n+2l\right)  !}}\left(  -1\right)
	^{l}\left(  4g^{2}\right)  ^{l}\exp\left(  -2g^{2}\right)  L_{n}^{2l}\left(
	4g^{2}\right)
\end{multline}
In this experiment the parameter $g$ is large, which means that the average
number of excitations  is also large $N\approx g^{2}$. This fact allows us to
use the following asymptotic expression for $L_{n}^{2l}\left(  4g^{2}\right)
$. With this approximation we arrive at the following expression for the maximum
of the cosine matrix element
\[
\underset{n,l\neq0}{\max}\left\vert \left\langle n\right\vert \cos\left(
2\sqrt{2}gx\right)  \left\vert n+2l\right\rangle \right\vert \approx\frac
{1}{\sqrt{2\pi g}}%
\]Hence, in the worst case we demand that
\begin{equation}
	\frac{V}{2}\frac{1}{\sqrt{2\pi g}}<<2,\label{perturbation}%
\end{equation}
then the condition (\ref{Perturbation_condition}) will be fulfilled automatically.
Or in ordinary units
\begin{equation}
	\frac{V}{2\omega}=\frac{\omega_{q}}{\omega}<<2\sqrt{2\pi\frac
		{g}{\omega}}\approx5\sqrt{\frac{g}{\omega}}.\label{condition_final}%
\end{equation}
For the experimental parameters of the here discussed experimental system (e.g. for the data of Fig. 4 of the main text: $\frac{\omega_{q}}{\omega}\leq2,$ $\ 2\sqrt{\frac{g}{\omega}}\approx\allowbreak11$) this condition is approximately fulfilled.\\
Upon introducing the displacement operator $D(\alpha)$, the last term in the Hamiltonian \ref{hamiltonian22} can be brought to a more pleasant form
\begin{multline}
	\cos\left(  2g\left(  a^{\dagger}+a\right)  \right)  = \\ \frac{1}{2}\left[
	\exp\left(  2ig\left(  a^{\dagger}+a\right)  \right)  +\exp\left(  -2ig\left(a^{\dagger}+a\right)\right)\right]\\
	=\frac{1}{2}\left[  D\left(  2ig\right)  +D\left(  -2ig\right)  \right].
	\label{Displ}
\end{multline}
By using the representation of the displacement-operator in Fock basis (see \cite{Glauber})
\[
\left\langle n\right\vert D\left(  \alpha\right)  \left\vert n\right\rangle
=\exp\left(  -\frac{\left\vert \alpha\right\vert ^{2}}{2}\right)  L_{n}\left(
\left\vert \alpha\right\vert ^{2}\right)
\]
where $L_{n}$ are the Laguerre polynomials. The eigenvalues and the solution of our problem can be written as
\[
E_{n}\approx n+\frac{V}{2}\exp\left(  -2g^{2}\right)  L_{n}\left(
4g^{2}\right)
\]%
\begin{equation}
	\left\vert \Psi\left(  t\right)  \right\rangle \approx\exp\left(  -it\left(
	A+D\right)  \right)
	{\displaystyle\sum\limits_{n=0}^{\infty}}
	a_{n}\left(  0\right)  \left\vert n\right\rangle , \label{Approx_Solution}%
\end{equation}
where $A$ and $D$ are diagonal matrices in the Fock space
\[
A_{nm}+D_{nm}=\left[  n+\frac{V}{2}\exp\left(  -2g^{2}\right)  L_{n}\left(
4g^{2}\right)  \right]  \delta_{nm}.
\]
We apply the solution (\ref{Approx_Solution}) when the state initially prepared as
\[
\left\vert \Psi\left(  0\right)  \right\rangle \rightarrow\mathcal{N}%
_{1}\left[  \left\vert 2k\right\rangle +\left\vert -2k\right\rangle
\right]  \rightarrow
\]%
\[
\mathcal{N}_{2}\left[  \left\vert ig\right\rangle +\left\vert -ig\right\rangle
\right]
\]
where $\mathcal{N}_1$ and $\mathcal{N}_2$ are normalization parameters and $\left\vert g\right\rangle $ is the coherent state with the displacement $g$. \\
\\
\onecolumngrid
After normalization it takes the following form%
\begin{equation}
	\left\vert \Psi\left(  0\right)  \right\rangle =\frac{1}{\sqrt{2}\sqrt
		{1+\exp\left(  -2g^{2}\right)  }}\exp\left(  -\frac{g^{2}}{2}\right)
	{\displaystyle\sum\limits_{n=0}^{\infty}}
	\frac{\left(  ig\right)  ^{n}+\left(  -ig\right)  ^{n}}{\sqrt{n!}}\left\vert
	n\right\rangle \label{initial-state}%
\end{equation}
then the state at time $t$ is%
\begin{equation}
	\left\vert \Psi\left(  t\right)  \right\rangle = \frac{1}{\sqrt{2}\sqrt
		{1+\exp\left(  -2g^{2}\right)  }}\exp\left(  -\frac{g^{2}}{2}\right)
	{\displaystyle\sum\limits_{n=0}^{\infty}}
	\left(  \frac{\left(  ig\right)  ^{n}+\left(  -ig\right)  ^{n}}{\sqrt{n!}%
	}\right)\exp\left(  -it\left(  n+\frac{V}{2}\exp\left(  -2g^{2}\right)
	L_{n}\left(  4g^{2}\right)  \right)  \right)  \left\vert n\right\rangle .
\end{equation}
The overlap with the initial state is equal%
\begin{equation}
	\left\langle \Psi\left(  0\right)  \right.  \left\vert \Psi\left(  t\right)
	\right\rangle =\frac{\exp\left(  -g^{2}\right)  }{2\left(  1+\exp\left(
		-2g^{2}\right)  \right)  }\left[
	{\displaystyle\sum\limits_{n=0}^{\infty}}
	\frac{\left[  \left(  ig\right)  ^{n}+\left(  -ig\right)  ^{n}\right]  ^{2}%
	}{n!}\exp\left(  -it\left(  n+\frac{V}{2}\exp\left(  -2g^{2}\right)
	L_{n}\left(  4g^{2}\right)  \right)  \right)  \right]  = \label{overlap}%
\end{equation}%
\[
\frac{\exp\left(  -g^{2}\right)  }{\left(  1+\exp\left(  -2g^{2}\right)
	\right)  }\left[
{\displaystyle\sum\limits_{n=0}^{\infty}}
\frac{g^{2n}\left(  1+\left(  -1\right)  ^{n}\right)  }{n!}\exp\left(
-it\left(  n+\frac{V}{2}\exp\left(  -2g^{2}\right)  L_{n}\left(
4g^{2}\right)  \right)  \right)  \right]  =
\]%
\[
\frac{1}{\cosh\left(  g^{2}\right)  }\left[
{\displaystyle\sum\limits_{n=0}^{\infty}}
\frac{g^{4n}}{\left(  2n\right)  !}\exp\left(  -it\left(  2n+\frac{V}{2}%
\exp\left(  -2g^{2}\right)  L_{2n}\left(  4g^{2}\right)  \right)  \right)
\right]  .
\]
\twocolumngrid
\subsubsection{Dynamics of the Mean Position $\left\langle x \right\rangle$}
\begin{figure}[H]
	\centering
	\includegraphics[width=1\linewidth]{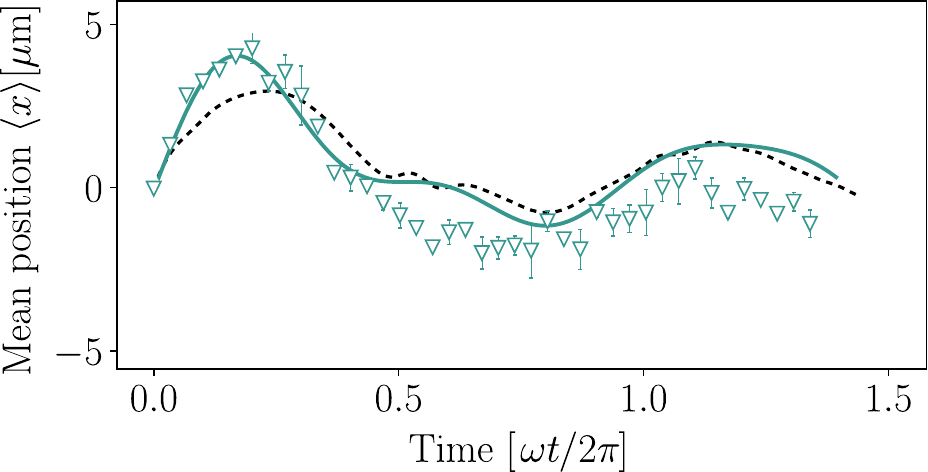}
	\caption{Temporal evolution of the measured values for the mean atomic position $\langle x\rangle$ for a qubit frequency of  $\omega_q/2\pi=\SI{1750(25)}{\hertz}$ (green data points), along with theory based on the perodic quantum Rabi model (dotted black lines) and the perturbative approach presented above. Experimental parameters are as in Fig. 3 of the main text.}
	\label{fig:S6}
\end{figure}

\vspace{-0.4cm}
In Fig. \ref{fig:S6} we have calculated the mean position coordinate $\left\langle x \right\rangle$ and can compare this observable directly to the numerical solution which was used in the main text and in the SI to calculate the theoretical predictions for the experimental data in Figure \ref{fig:S1}. We find good agreement of the perturbative method with the behavior of the mean position, which among other things is used to calculate the bosonic excitation number in Fig. 2 of the main text.
\subsubsection{Overlap $\left\vert \left\langle \Psi\left(  0\right)  \right.
	\left\vert \Psi\left(  t\right)  \right\rangle \right\vert ^{2}$}
\begin{figure}[H]
	\centering
	\includegraphics[width=1\linewidth]{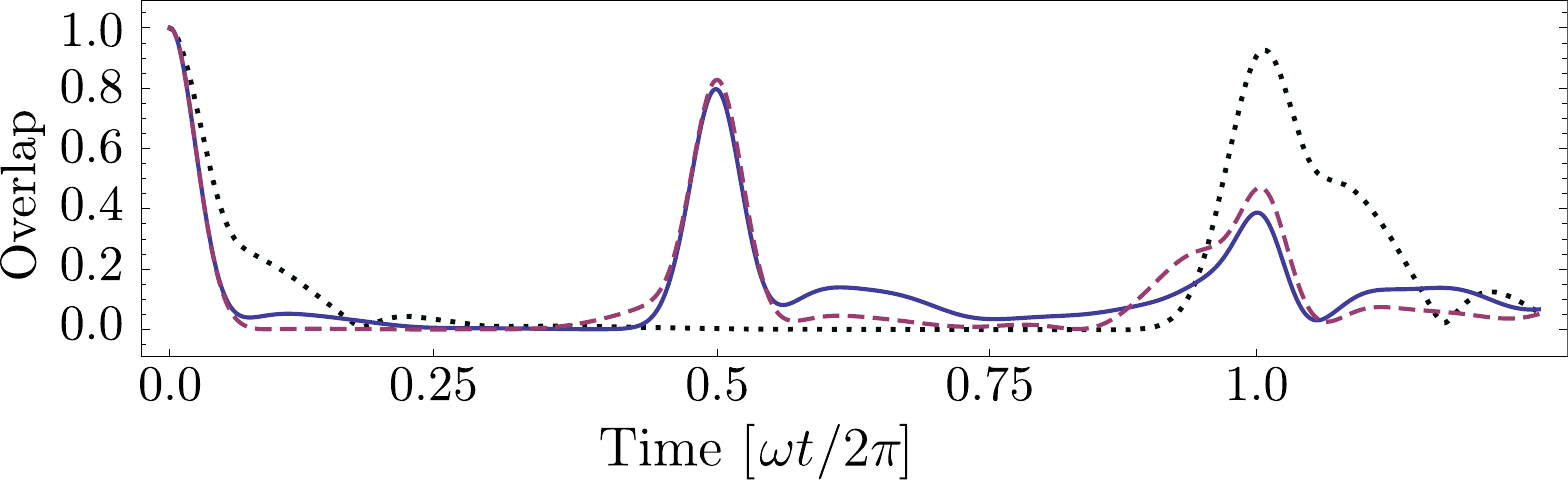}
	\caption{\justifying Variation of the overlap $\left\vert \left\langle \Psi\left(0\right)  \right.
		\left\vert \Psi\left(  t\right)  \right\rangle \right\vert ^{2}$ versus time (units of $\omega t/2\pi$). The initial state here is $\frac
		{1}{\sqrt{2}}\left[  \left\vert 2k\right\rangle +\left\vert -2k%
		\right\rangle \right]  $. The other parameters are $\frac{\omega_{q}}{\omega
		}=\frac{V}{2\omega}=\frac{1280}{650}\approx\allowbreak1.96\,$, $\frac
		{g}{\omega}=4.8$ the same as in Fig. 4 of the main text. The blue line corresponds the numerical calculations of eq. \eqref{schrodinger11}. The purple dashed line corresponds to the perturbative analytic expression eq. \eqref{overlap}. The black dotted line is the numerical calculation of the 'usual' QRM (see also Fig. 4), which as mentioned before, shows temporal dynamics at half of the periodicity of the pQRM (first revival at $\omega t/2\pi=1$).}
	\label{fig:S4}
\end{figure}

\begin{figure}[H]
	\centering
	\includegraphics[width=1\linewidth]{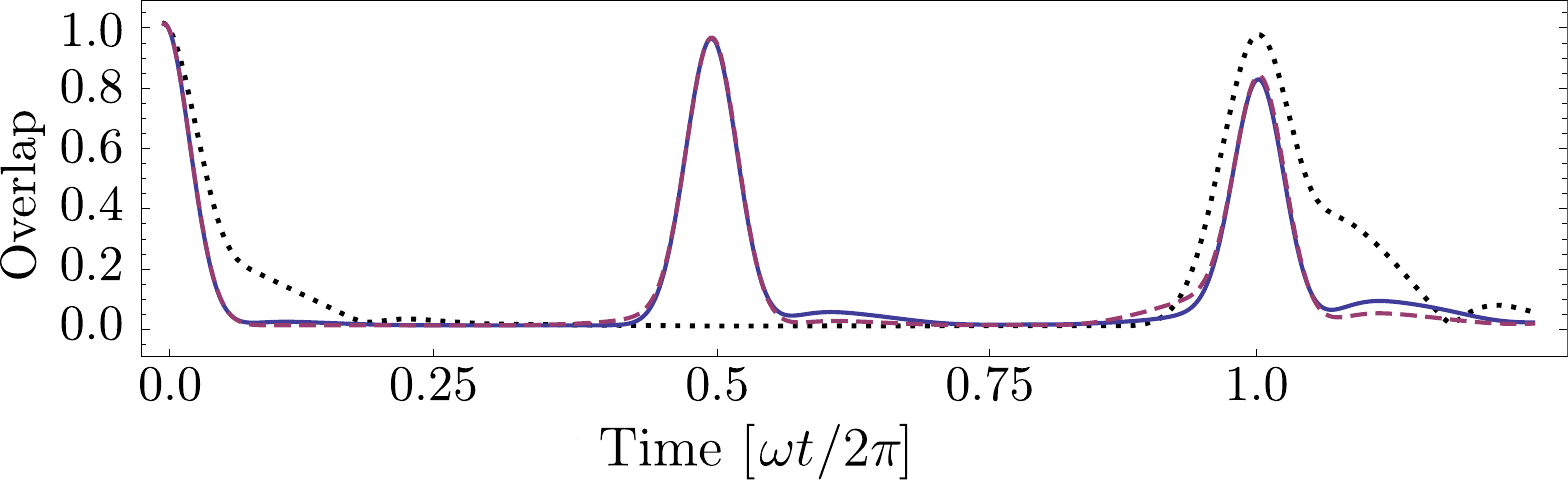}
	\caption{As in Fig. \ref{fig:S4}, but for smaller value of the qubit splitting $\frac{\omega_{q}}{\omega}%
		=1.2$.}
	\label{fig:S5}
\end{figure}

	\vspace{-0.4cm}
In Fig. \ref{fig:S4} and \ref{fig:S5} we compare the overlap $\left\vert \left\langle \Psi\left(  0\right)  \right.
\left\vert \Psi\left(  t\right)  \right\rangle \right\vert ^{2}$ obtained from
the numerical solution of eq. (\ref{schrodinger11}) for the initial state
(\ref{initial-state}) with the overlap according to eq. (\ref{overlap}).
A notable feature of this expression is that, due to the anharmonic nature of the
spectrum, the wave function does not recover after one oscillator period.
Moreover, it quite accurately predicts the amplitude of the overlap of the
wave function with the initial state. There is room to improve the accuracy of
this perturbative solution by taking into account higher order corrections.
However, these correction are outside of the scope of this note.
\subsubsection{Fidelity $\left\vert\left\langle \Psi_{_{\mathrm{p}}}\left(
	t\right)  \right.  \left\vert \Psi\left(  t\right)  \right\rangle \right\vert$}
	\vspace{-0.4cm}
In order to further justify the validity of this alternative method, we have calculated the fidelity of the exact and perturbative solutions
\[
F\left(  t\right)  =\left\vert \left\langle \Psi_{_{\mathrm{p}}}\left(
t\right)  \right.  \left\vert \Psi\left(  t\right)  \right\rangle \right\vert
\]
where  $\left\vert \Psi\left(  t\right)  \right\rangle $ is the exact solution
of the Schrödinger equation and   $\left\vert \Psi_{_{\mathrm{p}}}\left(  t\right)  \right\rangle $ is the perturbative one (see: Eq.
(\ref{Approx_Solution}))  with the initial state
\begin{equation}
	\left\vert \Psi\left(  0\right)  \right\rangle =\exp\left(  -\frac{g^{2}}%
	{2}\right)
	{\displaystyle\sum\limits_{n=0}^{\infty}}
	\frac{\left(  -ig\right)  ^{n}}{\sqrt{n!}}\left\vert n\right\rangle
\end{equation}

\begin{figure}[H]
	\vspace{2em}
	\centering
	\includegraphics[width=1\linewidth]{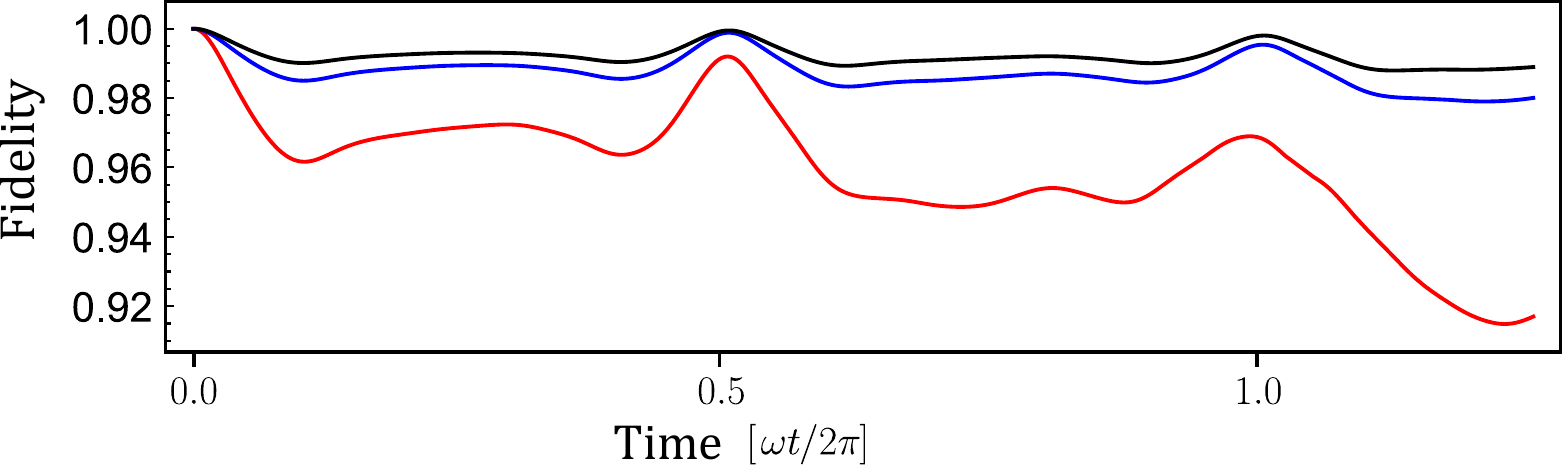}
	\caption{Temporal evolution of the fidelity $F\left(  t\right) =\left\vert\left\langle \Psi_{_{\mathrm{p}}}\left(
		t\right)  \right.  \left\vert \Psi\left(  t\right)  \right\rangle \right\vert$ for different values of the qubit frequency  $\omega_q/2\pi=\SI{650}{\hertz}$ (black), $\SI{800}{\hertz}$ (blue) and $\SI{1280}{\hertz}$ (red).}
	\label{fig:S7}
\end{figure} 
As can be seen in Fig. \ref{fig:S7}, the there is good agreement between the exact solution and the perturbative method. Increasing the qubit frequency leads to a larger deviation in the calculated value for $F(t)$, which is expected.

\bibliography{references}

\end{document}